\documentclass[12pt,preprint]{emulateapj}

\usepackage{natbib}
\newcommand{\Nsat}{N_\mathrm{sat}}

\newcommand{\hMpc}{h^{-1}\mathrm{Mpc}}
\newcommand{\hkpc}{h^{-1}\mathrm{kpc}}
\newcommand{\hMsun}{h^{-1}M_{\odot}}

\newcommand{\Mmin}{M_\mathrm{min}}
\newcommand{\Mzero}{M_\mathrm{0}}
\newcommand{\Mone}{M_\mathrm{1}}
\newcommand{\Mstar}{M_\ast}
\newcommand{\cgal}{c_\mathrm{gal}}
\newcommand{\fgal}{f_\mathrm{gal}}
\newcommand{\onehalo}{\mathrm{1halo}}
\newcommand{\twohalo}{\mathrm{2halo}}
\newcommand{\expon}{\mathrm{exp}}

\newcommand{\Rvir}{R_\mathrm{vir}}
\newcommand{\Lstar}{L^{\ast}}
\newcommand{\xir}{\xi(r)}
\newcommand{\xigg}{\xi_{\mathrm{gg}}}
\newcommand{\PNM}{P(N|M)}
\newcommand{\wprp}{w_{p}(r_{p})}
\newcommand{\chisqr}{\chi ^{2}}
\newcommand{\deltavir}{\Delta_\mathrm{vir}}
\usepackage{epsfig}
\usepackage{graphicx} 
\usepackage{rotating}

\begin{document}

  \title{THE EXTREME SMALL SCALES: DO SATELLITE GALAXIES TRACE DARK MATTER?}

\author{ Douglas~F.~Watson\altaffilmark{1},
  Andreas~A.~Berlind\altaffilmark{1,2},
  Cameron~K.~McBride\altaffilmark{1},
  David~W.~Hogg\altaffilmark{3,4}, 
  Tao~Jiang\altaffilmark{3}
  }

\altaffiltext{1}{Department of Physics and Astronomy, Vanderbilt
University, 1807 Station B, Nashville, TN 37235}
\altaffiltext{2}{Alfred P. Sloan Fellow} \altaffiltext{3}{Center for
Cosmology and Particle Physics, Department of Physics, New York
University, New York, NY, 10003} \altaffiltext{4}{Max-Planck-Institut
f$\ddot{\mathrm{u}}$r Astronomie, K$\ddot{\mathrm{o}}$nigstuhl 17,
D-69117 Heidelberg, Germany}

\begin{abstract}\label{abstract}

We investigate the radial distribution of galaxies within their host
dark matter halos  as measured in the Sloan Digital Sky Survey by
modeling their small-scale clustering.  Specifically, we model the
Jiang et al. (2011) measurements of the galaxy two-point correlation
function down to very small  projected separations ($10  \leq r \leq
400 \hkpc$), in a wide range of luminosity threshold samples (absolute
$r$-band magnitudes of $-18$ up to $-23$).  We use a halo occupation
distribution (HOD) framework with free  parameters that specify both
the number and spatial distribution of galaxies within their host dark
matter  halos.  We assume one galaxy resides in the halo center and additional galaxies are considered satellites that follow a radial
density profile similar to the dark matter Navarro-Frenk-White (NFW)
profile, except  that the concentration and inner slope are allowed to
vary.  We find that in low luminosity samples ($M_{r} < -19.5$  and
lower), satellite galaxies have radial profiles that are consistent
with NFW.  $M_{r} < -20$ and brighter satellite galaxies have  radial
profiles with significantly steeper inner slopes than NFW (we find
inner logarithmic slopes ranging from $-1.6$ to $-2.1$, as opposed to
$-1$ for NFW).  We define a useful metric of concentration,
$M_{1/10}$, which is the  fraction of satellite galaxies (or mass)
that are enclosed within one tenth of the virial radius of a halo.  We
find that $M_{1/10}$ for low luminosity satellite galaxies agrees with
NFW, whereas for luminous galaxies it is  $2.5-4$ times higher,
demonstrating that these galaxies are substantially more centrally
concentrated within their dark matter halos than the dark matter
itself.  Our results therefore suggest that the processes that govern
the spatial distribution of galaxies, once they have merged into
larger halos, must be luminosity dependent, such that luminous
galaxies become poor tracers of the underlying dark matter.

\end{abstract}



\section{INTRODUCTION}\label{intro}


Determining the relationship between the spatial distributions of galaxies 
and dark matter remains one of the central problems of theoretical and
observational cosmology. The radial distribution of galaxies within
their host dark matter halos dictates the correlation function on scales 
smaller than the virial radii of the largest dark matter halos
($\lesssim 1 \hMpc$). Therefore, the measured galaxy correlation
function itself is a powerful tool for shedding light on how galaxies 
trace the underlying dark matter within halos.

The halo occupation distribution (HOD) framework has been established
as a robust method for modeling galaxy clustering by simply
characterizing the biased relationship between galaxies and mass (see,
e.g., \citealt{peacock00a, scoccimarro01a, berlind02, cooray02,
zheng05}).  The HOD describes this relation by specifying the probability
distribution $\PNM$ that a halo of mass $M$ contains $N$ galaxies,
along with a prescription for the spatial distribution of galaxies 
within halos.  For this latter component, it is typicaly assumed that a 
single 'central' galaxy lives at the center of each halo, with additional 
'satellite' galaxies tracing the dominant dark matter component.  This
assumption has been used successfully to model galaxy clustering on scales 
larger than $r \sim 100 \hkpc$
\citep[e.g.,][]{zehavi04a,zehavi05a,zehavi11,zheng07,zheng09}. However,
\citet{watson_etal10} showed that this assumption does not work to explain
the smaller-scale clustering of luminous red galaxies (LRGs) in the
Sloan Digital Sky Survey (SDSS; \citealt{york00a}).  In order to achieve a 
good fit to the clustering of LRGs on scales smaller than $\sim 100 \hkpc$ 
\citep[measured by][]{masjedi06a}, they found that the radial profile of 
satellite LRGs must have a much steeper inner slope compared to the 
Navarro-Frenk-White (NFW; \citealt{nfw97}) dark matter profile.  
\citet{watson_etal10} concluded that the distribution of satellite LRGs 
within halos is better described by an isothermal distribution.

Recently, \citet{jiang_etal11a} measured the projected two-point
correlation function $\wprp$ for several luminosity samples from the SDSS, 
extending the measurements done by \citet{zehavi11} on intermediate scales 
($\sim 0.1 - 40 \hMpc$) down to extremely small galaxy-galaxy
separations ($0.01 - 7 \hMpc$).  Our motivation for this paper is to
model the innermost data points ($0.01 - 0.4 \hMpc$) in the same vein
as \citet[][hereafter W10]{watson_etal10} to see if there is a
luminosity dependence of the radial profile of satellite galaxies.

The paper is laid out as follows.  In \S\ref{data}, we review the
\citet[][hereafter J11]{jiang_etal11a} measurements.  In \S\ref{method}, 
we discuss our modeling method in the following manner: in
\S\ref{theory} we provide an overview of the general technique used to 
model the small-scale correlation function; in \S\ref{the_models} 
we revisit the W10 four-parameter PNM model that allows the the probability 
distribution $\PNM$ to vary, and the five-parameter PNMCG model that also 
allows the radial distribution of galaxies within halos to vary.
We present our results in \S\ref{results}.  Finally, we summarize and
discuss the implications of our results in \S\ref{discussion}.
Throughout the paper, we assume a $\Lambda$CDM csomology with
$\Omega_{m}=0.25, \Omega_{\Lambda}=0.75, \Omega_{b}=0.04, h_{0}=0.7,
\sigma_{8}=0.8, n_{s}=1.0$.
   

\section{DATA}\label{data}


J11 measured the projected two-point correlation function down to very
small scales ($0.01 < r < 7 \hMpc$) for a large range of
volume-limited galaxy luminosity threshold samples.  Galaxy number
denisities and median redshifts for each sample can be found in
Table~\ref{model_table}.  These samples are constructed from the NYU
VAGC \citep{VAGC_05} V7.2 data \citep{DR7_09} which contains
$8.6\times10^{5}$ SDSS Main Sample galaxies \citep{strauss02}.
Spectra for each luminosity threshold sample are cross-correlated with
the full imaging sample to compute $\wprp$ (details of the method can
be seen below).  The full imaging sample consists of $\sim10^{8}$
galaxies drawn from the SDSS imaging catalog.

Measuring $\wprp$ on very small scales is a non-trivial task.  The
SDSS spectroscopic sample suffers from incompleteness due to the
physical size of the fiber-optic cables, which impede the ability to
take spectra of two galaxies closer than $55''$ on the sky.  These
``fiber collisions'' thus result in a minimum pair separation of $55''$.
This effect is partially mitigated by the tiling method \citep{blanton03},
which overlaps spectroscopic plates in order to yield full sky coverage.
In plate overlap regions, galaxy pairs closer than $55''$ can be recovered.
However, this still leaves $\sim 9\%$ of galaxies without measured
redshifts (J11) and this can strongly affect clustering measurements on
very small scales.  Following the approach described in \citet{masjedi06a,masjedi08}, 
J11 used a cross-correlation technique between the imaging and spectroscopic 
samples to correct for fiber collisions and obtain unbiased $\wprp$ 
measurements down to the $10\hkpc$ scale.  Details of the method can be 
found in \S\ref{method} of J11.

A second potential problem that affects very small scales is that galaxies 
may be overlapping and the light within this region needs to be properly
distributed between the two galaxies.  This is known as ``deblending''.  
M06 found a systematic error in the SDSS pipeline wherein too much light 
was allocated to the dimmer of the two galaxies.  As a result, a galaxy 
that may have been too dim to make the LRG brightness cut could now be 
included in the sample.  This increased the number of small-scale pairs 
and boosted the correlation function on very small scales.  M06 quantified
this effect and corrected their measurements accordingly.  Since the physical 
sizes of galaxies decrease rapidly with decreasing luminosity, this photometric
deblending error diminishes in lower luminosity samples.  For this reason,
J11 ignored this effect in the luminosity samples they considered
($M_{r} < -18$ through $-21$).  Nevertheless, we estimate the maximum effect of 
deblending by applying the M06 LRG correction to the $M_{r} < -21$ sample and
repeating our analysis.  We find that our results do not change qualitatively.

We restrict ourselves to modeling the J11 data points from $\sim 10 - 400
\hkpc$.   These can be seen in Figure~\ref{fig:wpgg_ALL}, which shows
$\wprp$ scaled by an $r_{p}^{-1}$ power law, with each panel corresponding
to a distinct luminosity threshold.  The bottom right panel of 
Figure~\ref{fig:wpgg_ALL} shows the LRG data points measured by M06 and 
modeled by W10.  Along with these 9 data points used in our modeling, we also
incorporate the measured number density for each luminosity sample
(see Table~\ref{model_table}), providing a tenth data point.  We use
the full covariance matrices from the 9 J11 $\wprp$ data points for each
sample, which were estimated using jackknife resampling of 50 regions
on the sky.  The same jackknife samples were used to estimate the error on 
each calculated number density.
      

\section{REVIEW OF THE METHOD}\label{method}



\subsection{The HOD and the Galaxy 2PCF}\label{theory}


Here we briefly review the method used in W10, which is based on the
halo occupation distribution (HOD) formalism.  The HOD fully
characterizes the number, velocity and spatial distribution of
galaxies within dark matter halos.  The probability distribution that
a virialized dark matter halo of mass $M$ will host $N$ galaxies is
designated as $\PNM$.  $\langle N \rangle _{M}$ is the first moment of
$\PNM$, and is the mean number of galaxies as a function of halo mass.
Motivated by theory \citep[e.g.,][]{berlind03,kravtsov04a,zheng05}, we
consider central galaxies that live at the center of their host halo
and satellite galaxies that orbit within the host potential as
separate terms.  We thus write the first moment of the HOD as 
$\langle N \rangle _{M} = 1 + \langle \Nsat \rangle _{M}$.  Furthermore,
we assume that there is some minimum halo mass, $\Mmin$, below which a 
halo will contain no galaxies and above which there will always be at 
least one central galaxy.  For the satellite component, we adopt the
parametric form, $\langle \Nsat \rangle _{M} =
\expon[-{\Mzero}/(M_{\mathrm{halo}}-\Mmin)]\times(M_{\mathrm{halo}}/\Mone)^{\alpha}$,
where the satellite galaxies obey a power-law function of slope
$\alpha$ with an exponential cut-off at the low mass end at $\Mzero$.
$\Mone$ is the characteristic mass scale where a halo will contain, on
average, one central and one satellite galaxy.  

To calculate the mean number of satellite galaxy pairs $\langle \Nsat(\Nsat-1)\rangle$,the second moment of the satellite $\PNM$, we assume that the number 
of satellite galaxies in a halo of mass $M$ follows a Poisson distribution 
of mean $\langle \Nsat \rangle_{M}$.  This sort of HOD
parameterization is widely used to model galaxy clustering data
\citep{zehavi05b,zehavi05a,zehavi11,tinker05,conroy06,zheng07,zheng09,watson_powerlaw11}.

Our fiducial model for characterizing the spatial distribution of galaxies within their host halos places one galaxy at the center and
assuming that the satellites trace the underlying dark matter density distribution.  
The dark matter density profiles are described by the NFW relation
$\rho(r)\propto(c\frac{r}{R_{vir}})^{-1}(1+c\frac{r}{R_{vir}})^{-2}$
\citep{nfw97}, where $c$ is the concentration parameter, $c\equiv
\Rvir/r_s$, and $r_{s}$ is the characteristic scale radius.  We use
the virial definition of a halo to calculate the virial radius of host
halos, such that $\Rvir = ((3M)/(4\pi \deltavir \bar{\rho}))^{1/3}$,
where $\deltavir$ is the mean halo overdensity ($\deltavir = 200$), and
$\bar{\rho}$ is the mean density of the universe.  The concentration -
mass relation is given by \citet{zheng07} for the modification of
\citet{bullock01}: $c=\frac{c_{0}}{(1 +
z)}\times(\frac{M_\mathrm{halo}}{M_{\ast}})^{-\beta}$, where
$c_{0}=11$, $M_{\ast}$ is the non-linear collapse mass at the median
redshift of the sample for our choice of cosmology ($\Mstar$ is
redshift dependent and is thus uniquely defined for each luminosity
sample and is given in Table \ref{model_table}) and $\beta =
0.13$. 



\begin{deluxetable}{cccc}
\tablecaption{INPUT DATA FOR ALL LUMINOSITY SAMPLES}
\tablewidth{.45\textwidth} \startdata \hline 	\hline \\[-2ex]
$M_{r}$  &  $\bar{n}_{g}$    &   $z_{med}$  & log($\Mstar$)  \\ \hline
\\[-2ex] $-18.0$     &  3.209 (0.169)  &  0.032  &  12.4228  \\
$-18.5$     &  2.405 (0.131)  &  0.039  &  12.4147  \\ $-19.0$     &
1.689 (0.092)  &  0.046  &  12.4057  \\ $-19.5$     &  1.326 (0.058) &
0.059  &  12.3887  \\ $-20.0$     &  0.749 (0.018)  &  0.078  &
12.3647 \\ $-20.5$     &  0.377 (0.009)  &  0.097  &  12.3397  \\
$-21.0$     &  0.123 (0.002)  &  0.116  &  12.3057  \\[-2ex] \enddata
\tablecomments{$\bar{n}_{g}$ is measured in units of $10^{-2} h^{3}
\mathrm{Mpc}^{-3}$ and the associated error from 50 jackknife samples
on the sky is shown in parentheses.  $z_{med}$ is the median redshift as
measured in J11.  $\Mstar$ is the non-linear collapse mass and is in
units of $\hMsun$.\\}
\label{model_table}
\end{deluxetable}


The mean number density of galaxies can be calculated for a given HOD 
by weighting the abundance of halos by $\langle N \rangle _{M}$ and 
integrating over all halo mass (see Eq.[2] of W10).  We adopt the 
\citet{warren06} halo mass function, $dn/dM$, in all of our calculations,
but our results are not sensitive to the specific choice of mass function.

To model the galaxy two-point correlation function (2PCF), $\xigg$, we
use the halo model.  $\xigg$ can be decomposed into contributions due
to galaxies residing in the same halo (the ``one-halo'' term) as well
as galaxies living in separate, distinct halos (the ``two-halo''
term).  Therefore, $\xigg$ can be written as the sum of the one-halo
and two-halo terms (e.g., \citealt{cooray02}; for this particular form
of the equation see \citealt{zheng04a}):
$\xi_{\mathrm{gg}}(r)=\xi_{\mathrm{gg}}^{\onehalo}+\xi_{\mathrm{gg}}^{\twohalo}
+ 1 $.  Since we are probing such small scales we need only consider
the one-halo term for modeling $\xi$.  However, it is possible that
omitting the two-halo term entirely could cause a bias in the best-fit
parameters for the one-halo term.  This bias would be largest for the
dimmest galaxy sample, because these galaxies live in the smallest
halos, and the two-halo term will become important at small scales.
However, even for these galaxies the effect is limited.  For instance,
while it is true that isolated $-18$ galaxies will reside in halos
with radii much smaller than $0.4 \hMpc$, the relevant halo size is
set by $\Mone$.  Furthermore, this is a threshold sample, so although
the $-18$ sample will be dominated by galaxies of that brightness,
there will still be much brighter galaxies that live in larger halos
that influence $\Mone$.  As seen in the upper-left panel of
Fig. \ref{fig:rad_dist}, $\langle \Mone \rangle$ for the $-18$ sample
is $10^{12.77}$.  This corresponds to a virial radius of order $0.4
\hMpc$.  To carefully test the influence of the two-halo term, we have
have measured where the 1- and 2-halo terms cross for the $-18$
sample.  This occurs at roughly $1\hMpc$, so the influence of the
2-halo term at $0.4 \hMpc$ has a small effect on $\wprp$.  We find
that the amplitude of the 2-halo term is $\sim 7\%$ of the one-halo
term at $0.4 \hMpc$ for the $-18$ sample.  Nevertheless, we do
approximately account for the effect of the 2-halo term on our
modeling, and we discuss these assumptions in detail below.

The one-halo term depends on the second moment of $\PNM$, as well as
the pair separation distributions of central-satellite and
satellite-satellite pairs (see Eq.[4] of W10 for the particular form).
The pair separation distribution for central-satellite pairs is
essentially the same as the density profile of satellite galaxies.
The satellite-satellite pair distribution is the convolution of the
density profile with itself and, as in W10, we use the \citet{sheth01}
calculation for the convolution of a truncated NFW profile with
itself.   Therefore, we can start to see the link between the shape of
$\xir$ and the shape of the satellite profile on small scales.  This
is discussed in detail in \S\ref{results}.

For each luminosity threshold, J11 measured the \emph{projected}
correlation function, so we convert our theoretical real-space
correlation function $\xir$ to $\wprp$ (Eq.[5] of W10).  At each
luminosity threshold, we integrate up to the $\pi_{\mathrm{max}}$
value given in \citet{zehavi11}, which ranges from $40-60\hMpc$.   The
method of J11 results in a $\pi_{\mathrm{max}}$ that is effectively
larger than our choice.  However, the difference is insignificant at
the small scales we model. Integrating out to large scales means that
the two-halo term cannot be entirely ignored.  Since calculating the
two-halo term correctly is fairly complex and its contribution to
$\wprp$ is minimal at the scales of the data points that we model, we
use a simple approximation instead.  We use a two-halo term whose
shape is the same as \citet{zehavi11}, but whose amplitude can change
with HOD parameters.  Thus, for a given set of HOD parameters, we
first calculate the large-scale bias of galaxies, $b_{g}$,
\begin{equation}
b_{g} =
\bar{n}_{g}^{-1}\int_{\Mmin}^{\infty}dM\frac{dn}{dM}b_{h}(M)\langle N
\rangle _{M} ,
\end{equation}
where $b_{h}(M)$ is the large-scale bias of halos from
\citet{tinker08a}.  The two-halo term has a simple relation to the
matter-matter correlation function on large scales:
$\xi_{\mathrm{gg}}^{\twohalo} = b_{g}^{2}\xi_{mm}$.  Thus, the relative two-halo
terms of two different galaxy samples really only depends on the ratio
of the bias of each sample. We consider the best-fit two-halo term
from \citet{zehavi11} for each luminosity and append a newly defined
two-halo term to our model one-halo term,
\begin{equation}
\xi_{\mathrm{gg}}^{\twohalo} = \Big(\frac{b_{g}}{b_{g}^{\
\prime}}\Big)^{2}\xi_{\mathrm{gg}}^{\twohalo \ \prime} ,
\end{equation}
where primes designate the \citet{zehavi11} values.  In effect, we
allow the amplitude of the 2-halo term to vary with HOD parameters,
but we keep its shape fixed.  The error introduced by this
approximation is completely negligible as we have checked that even a
shift as large as $20\%$ in the amplitude of the two-halo term has no
appreciable effect on $\wprp$ at the maximum pair separation that we
consider ($0.4\hMpc$) for any of the luminosity threshold  samples.
This test also emphasizes our insensitivity to the difference between
our choice of integrating out to the $\pi_{\mathrm{max}}$ given by
\citet{zehavi11} and the technique used by J11.


\begin{figure*}
\begin{center}
\includegraphics[angle=90]{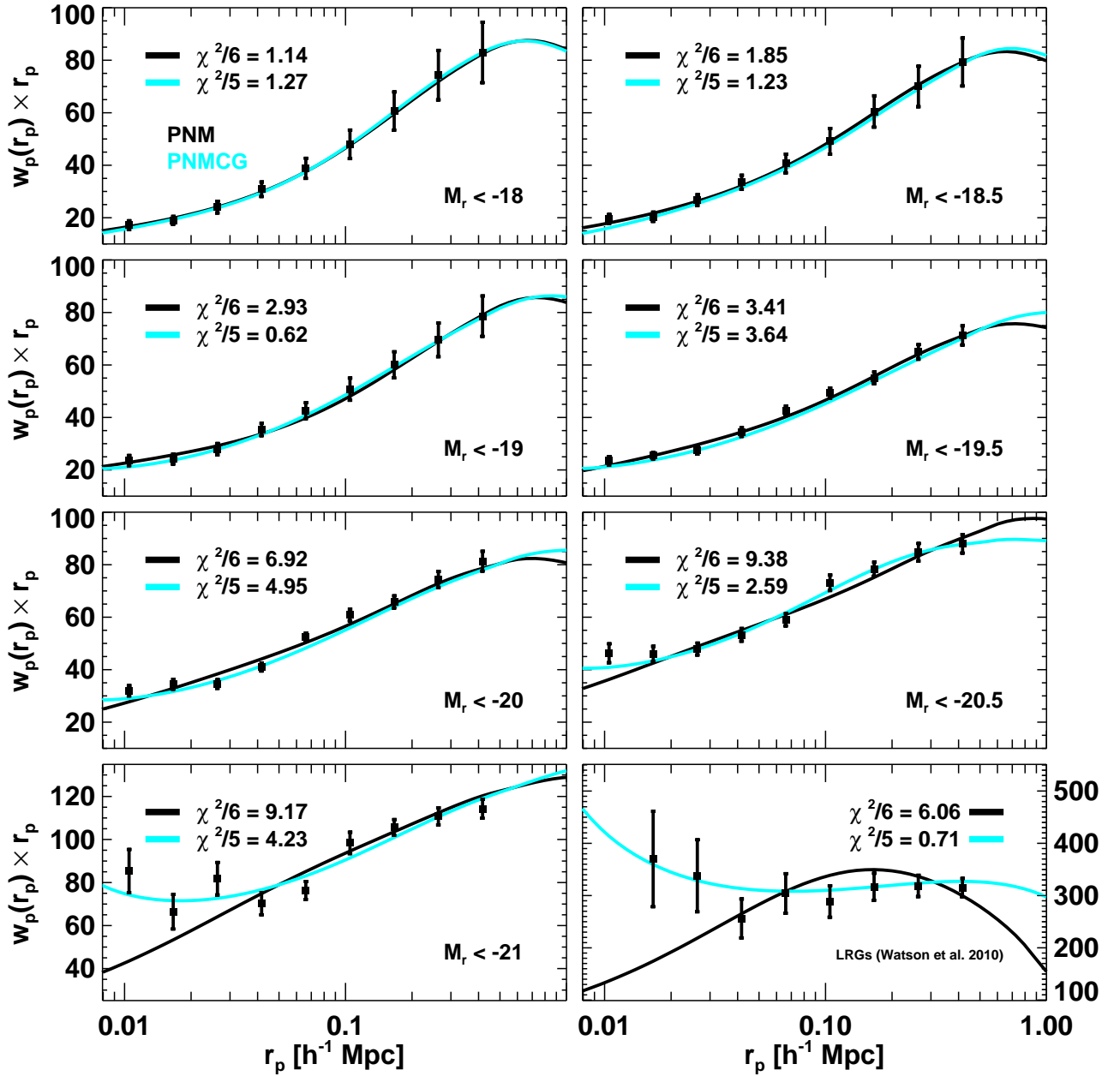}
\caption{Model fits to the projected correlation function for
luminosity thresholds spanning $M_{r} < -18$ to LRGs.  The points in
the $M_{r} < -18 \text{ through } M_{r} < -21$ panels show the Jiang
et al. (2011) $\wprp$ measurements (multiplied by $r_{p}$) and their
associated jackknife errors.  The bottom right panel shows results for
LRGs from \citet{watson_etal10}.  The black curves show the best-fit
PNM models, which use four-free parameters describing the probability
distribution $\PNM$ ($\Mmin,  \Mone, \Mzero$,and $\alpha$).  The
PNMCG model, shown in cyan, considers two additional free parameters
(but $\Mzero$ is fixed to $\Mmin$): (1) $\fgal$, which relates the
concentration of the density profile of satellite galaxies relative to
dark matter ($\fgal = \cgal/c$), and (2) $\gamma$, which allows for the
inner slope of the density profile to differ from the dark matter
distribution.  The reduced $\chi^2$ values  for each best-fit model
are listed in each panel.}
\label{fig:wpgg_ALL}
\end{center}
\end{figure*}



\begin{figure}
\includegraphics[width=.5\textwidth,height=.35\textheight]{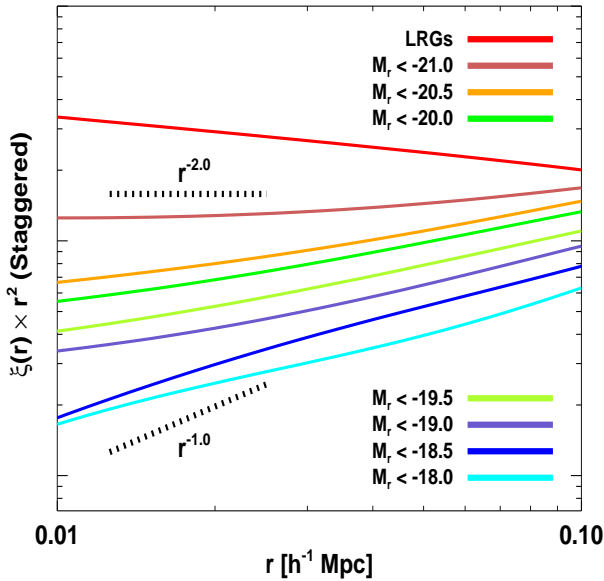}
\caption{The real-space correlation function $\xir$ residuals from an
$r^{-2}$ power law from the PNMCG best-fit models before converting to
the projected correlation functions of Fig.~\ref{fig:wpgg_ALL}. 
Amplitudes have been arbitrarily shifted for
clarity.  The slope of $\xir$ on small scales is a reflection of the
central-satellite pair distribution, which is essentially just the
density profile itself.  There is a strong luminosity dependence of
the slope of $\xir$ on small scales, becoming steeper and steeper for
the brighter galaxy samples and this carries directly over to the
luminosity dependence of the radial profile of satellite galaxies.
The dotted lines show $r^{-1}$ and $r^{-2}$ power-law slopes.}
\label{fig:xi_ALL}
\end{figure}



\begin{figure*}
\begin{center}
\includegraphics[width=.8\textwidth,height=.8\textheight,angle=0]{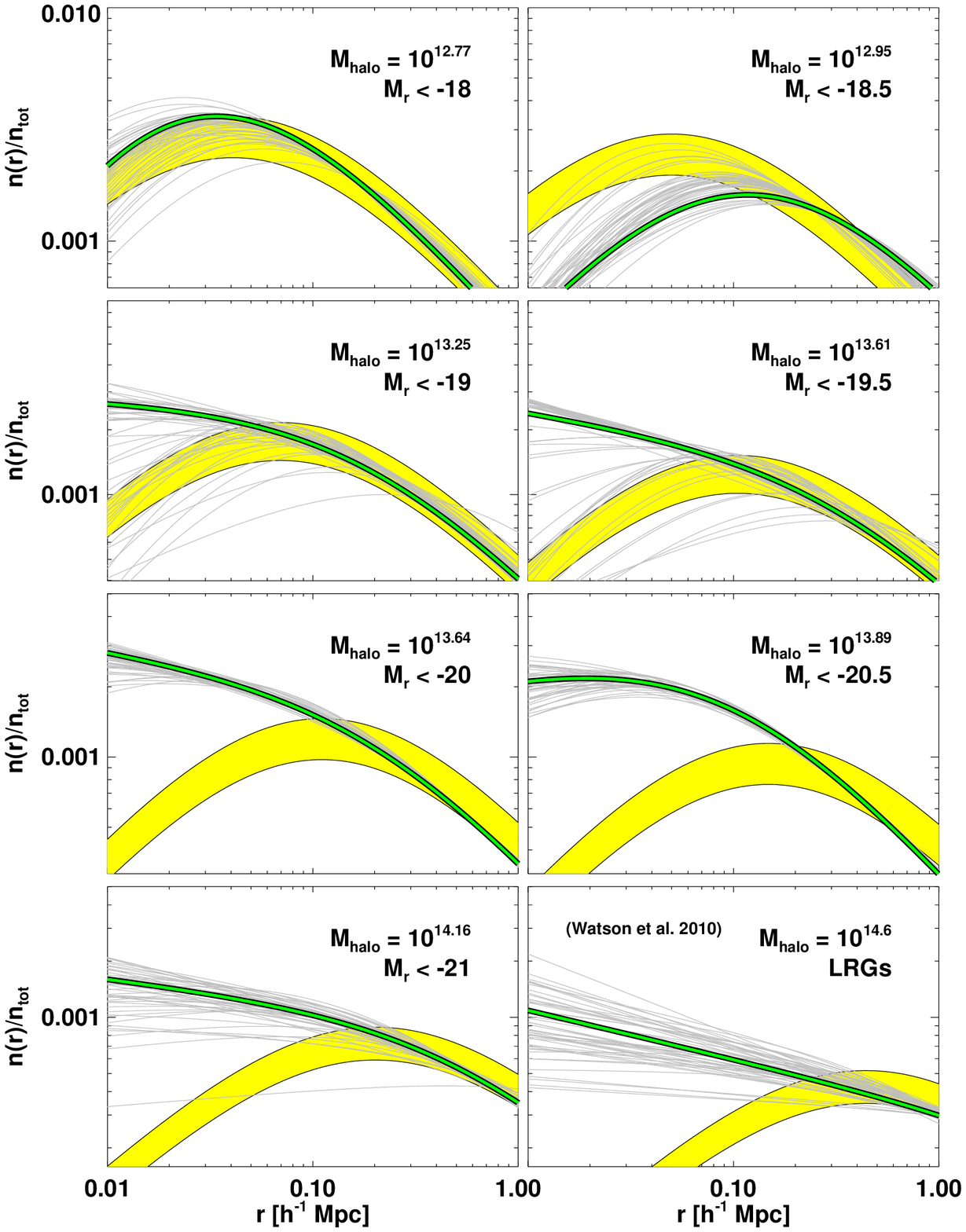}
\caption{The radial profile of satellite galaxies as a function of
luminosity.  Each panel shows results for a specific luminosity sample, and
for a halo mass that is chosen to be the mean value of $\Mone$ from the PNMCG
MCMC chain (listed in the top right of the panel).  The solid, yellow bands 
show the NFW profile, with a $20\%$ uncertainty to account for the possible 
ambiguity in the dark matter distribution.  Green curves show the profile
corresponding to the best-fit PNMCG model.  Each panel also contains 50 grey 
curves representing 50 randomly drawn links from the top $95\%$ of the 
PNMCG MCMC chain, after sorting in $\chi^{2}$ from lowest to highest values.
These 50 curves thus span the $95\%$ confidence region allowed by the data.
As we probe higher luminosities, the radial profile of satellite galaxies
strongly deviates from an NFW distribution on small scales, showing
that luminous galaxies are poor tracers of the underlying dark matter.}
\label{fig:rad_dist}
\end{center}
\end{figure*}



\subsection{The PNM and PNMCG Models}\label{the_models}


W10 considered 4 distinct models for modeling the small-scale $\wprp$
of LRGs.  In this paper we need only consider two of the models,
defined as:

\begin{itemize}
\item \textbf{\emph{PNM}} - in this model the only free parameters are
those associated with the $\PNM$ distribution as described in \S\ref{theory} 
- $\Mmin,\Mzero,\Mone, \text{ and }\alpha$.  While these
parameters are free to take on any value, the spatial distribution of
satellite galaxies within halos is fixed to an NFW dark matter density
profile.  Since there are 10 data points (9 from the J11 $\wprp$
measurements along with the measured number density for each
luminosity sample) and 4 free parameters, the PNM model has 6 degrees of
freedom.
\item \textbf{\emph{PNMCG}} - this model allows for the same free
parameters as the PNM model, with the exception of $\Mzero$. $\Mzero$
was unconstrained in our intitial MCMC runs (this was also the case in
W10), thus we fixed $\Mzero$ to $\Mmin$ throughout our analysis.  The
PNMCG model also considers a parametrized density profile for
satellite galaxies.  As in W10, we allow both the concentration
and the inner slope of the density profile to be free.  The 
satellite galaxy concentration can differ from the dark matter concentration
(defined in \S\ref{theory}) through the free parameter $\fgal$,
\begin{equation}\label{eq:conc}
\cgal =\fgal\times c .
\end{equation}

The inner slope of the density profile is no longer fixed to -1, as
is the case for an NFW profile, but rather is specfied by the free parameter 
$\gamma$.  As in W10, we adopt the following density profile for
satellite galaxies
\begin{equation}\label{eq:profile_gamma}
\rho(r) = \frac{\rho_{s}}{(\cgal \frac{r}{R_{vir}} )^{\gamma} (1+\cgal
\frac{r}{R_{vir}})^{3 - \gamma}} .
\end{equation}
For an NFW profile $\gamma = 1$, however, the PNMCG model allows
$\gamma$ to take on any value from $0-4$.  This model considers the
same number of data points as the PNM model, but now there are 5 free
parameters ($\Mmin,  \Mone, \alpha, \fgal$ and $\gamma$) resulting in 
5 degrees of freedom.
\end{itemize}

As detailed in \S3.4 of W10, we use a Markov Chain Monte Carlo (MCMC)
method to probe the parameter space for a given set of parameters (see
\citealt{dunkley05} for details on MCMC techniques).  When a chain has
converged, we can find the most likely value for each parameter by
calculating the mean of its distribution.  Errors for each parameter
are given by the extrema of the middle $68.3\%$ of the distribution.
Best-fit parameters are found by the combination of parameter values
for which $\chi^{2}$ is a minimum.


\section{RESULTS}\label{results}



\begin{figure}
\begin{center}
\includegraphics[width=.5\textwidth,height=.6\textheight,angle=0]{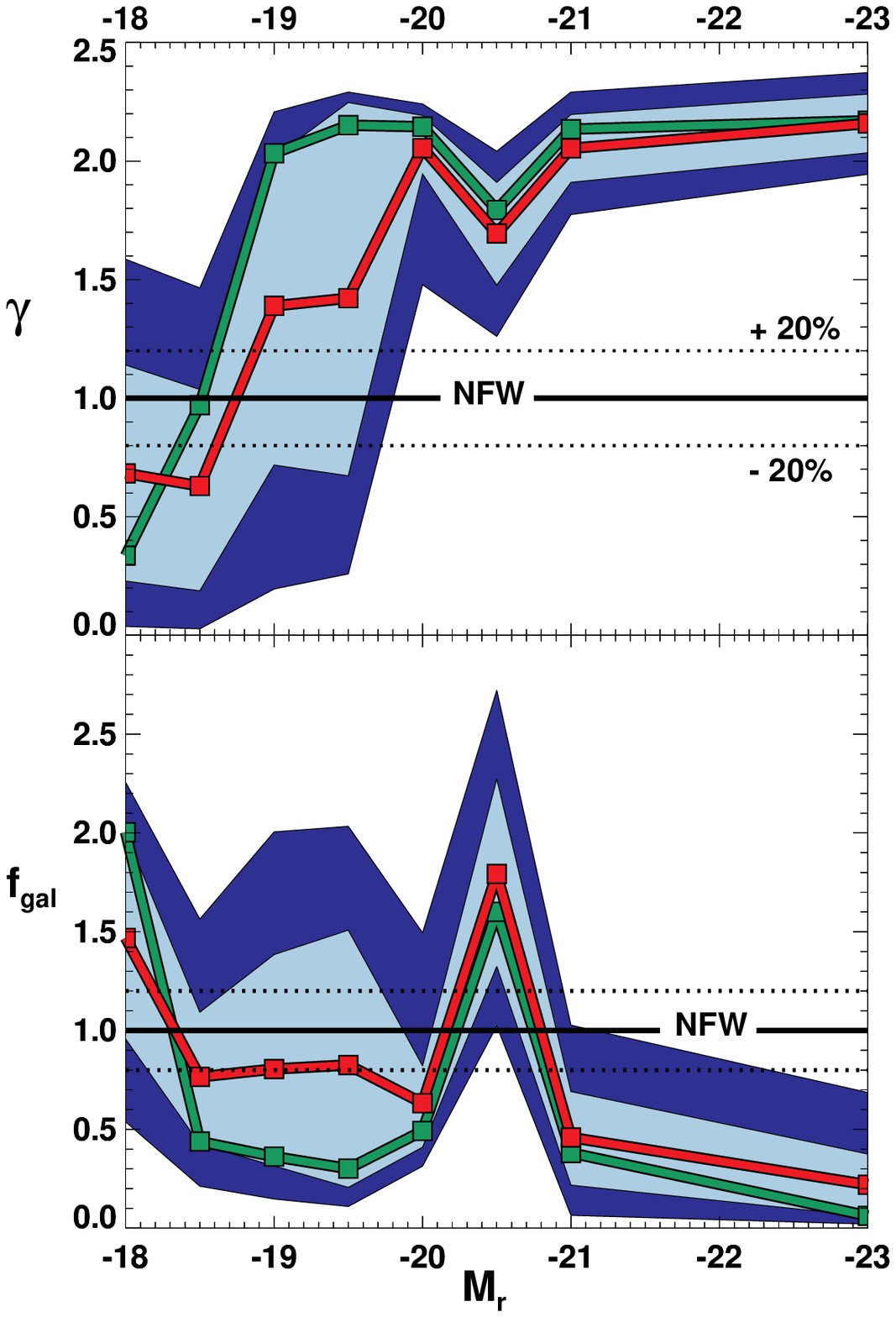}
\caption{\emph{Top Panel}: The slope of the inner density profile of
satellite galaxies as a function of galaxy luminosity.  The filled red
squares (and connecting lines) show the mean value of $\gamma$ from
the MCMC chain of the PNMCG model for  each luminosity sample and the
green squares represent the best fit.  The light blue and dark blue
bands are the associated errors from the extrema of the middle
$68.3\%$ and $95\%$ of the distribution, respectively.  $\gamma = 1$
corresponds to an inner slope for an NFW dark matter profile, and is
shown as a solid black line.  The dotted black lines highlight an
assumed $20\%$ inaccuracy in the dark matter profile.   \emph{Bottom
Panel}: The same procedure as the top panel, but for the parameter
$\fgal$ which relates the galaxy and dark matter concentrations
($\fgal = \cgal/c$). $\fgal$ and $\gamma$ are intrinsically linked and
there is a strong trend towards inner slopes becoming steeper than NFW
as we go to higher luminosities, with corresponding decreasing values
of $\fgal$.}
\label{fig:Mr_vs_Gamma}
\end{center}
\end{figure}



\begin{figure*}
\begin{center}
\includegraphics[width=.8\textwidth,height=.8\textheight,angle=0]{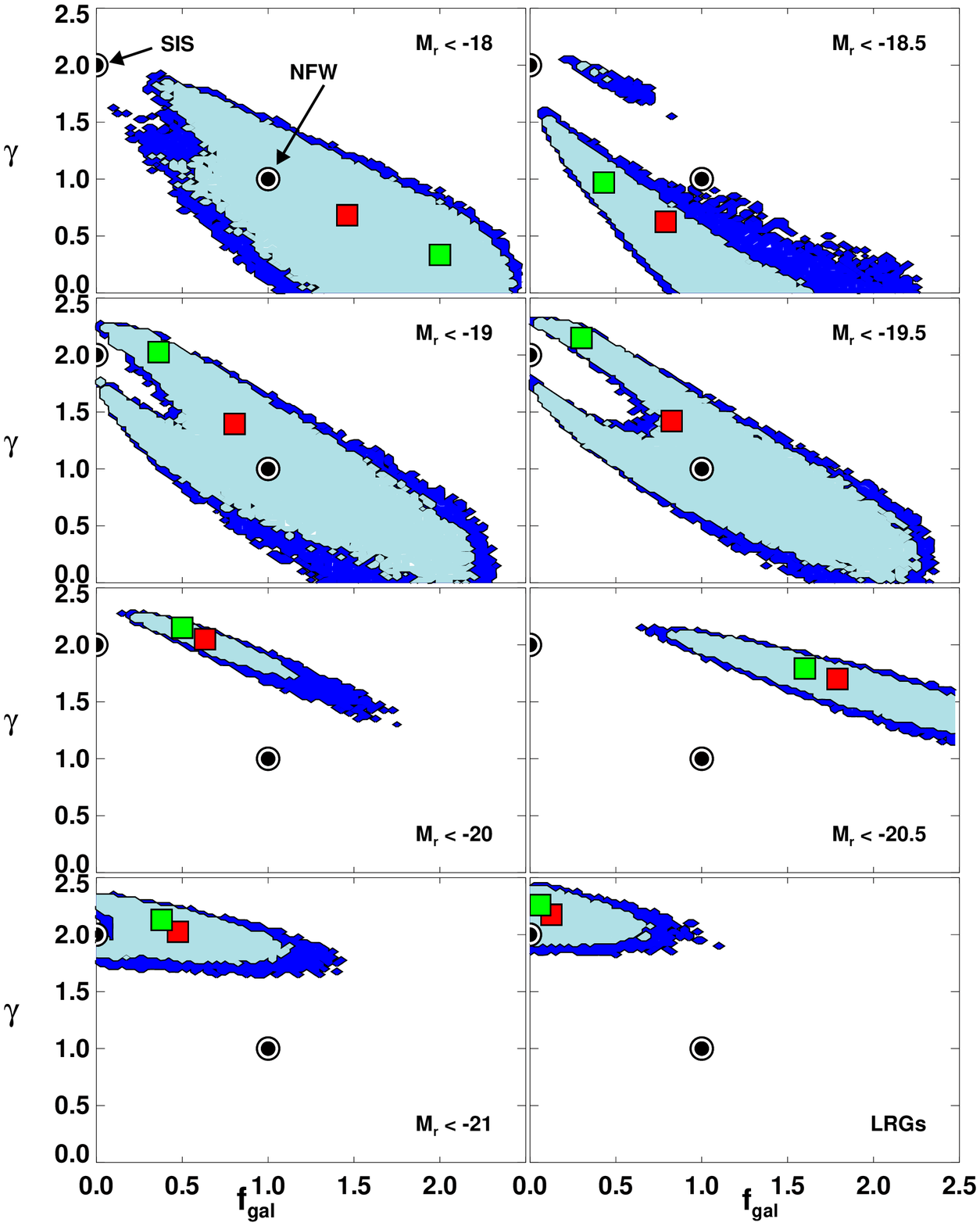}
\caption{The $\gamma - \fgal$ parameter space as a function of
luminosity, highlighting the relationship between the inner slope and
concentration of the density profile of satellite galaxies
relative to dark matter.  For each panel, the dark blue and light blue
regions are defined as the top $95\%$ and $68.3\%$ of the MCMC chain
for a given luminosity sample after sorting in $\chisqr$.  There are
two points of reference designated as filled circles - the $\gamma =
1, \fgal = 1$ combination that represents an NFW distribution, and
the $\gamma = 2, \fgal \rightarrow 0$ combination that represents a singular 
isothermal sphere distribution (SIS).  Red squares show the mean values
of $\gamma$ and $\fgal$ from our MCMC runs and green squares show the
values for the best-fit model.  The data for $M_{r} < -20$ and brighter
galaxies strongly rule out the NFW model and are in good agreement
with an SIS distribution for $M_{r} < -21$ and LRGs.}
\label{fig:gamma_fgal_space}
\end{center}
\end{figure*}


J11 measured $\wprp$ to very small scales over a large range in
luminosity thresholds.  Their measurements nicely overlap those of
\citet{zehavi11} on intermediate scales ($\sim 0.2 - 7 \hMpc$, see
Fig.~14 in J11), and extend down to very small scales with the
innermost data point for each sample at $\sim 10 \hkpc$.  As discussed
in \S\ref{the_models}, for each luminosity threshold sample, we model
$\wprp$ with, (1) the PNM  model, which only varies parameters that
determine the number of galaxies  in a given halo, but forces the
satellite galaxies to have an NFW spatial distribution  within their
halo, and (2) the PNMCG model, which also allows the spatial
distribution of satellite galaxies to vary within halos.

Figure 1 shows our modeling results for each luminosity sample. $\wprp$
has been scaled by an $r_{p}^{-1}$ power law to more clearly highlight any
discrepancies between the PNM and PNMCG models. Each  panel shows the
SDSS data points as well as the best-fit model for the PNM (black
curve) and PNMCG  (cyan curve) cases. It is clear from the figure that
as we go to higher luminosities, the PNMCG model  provides a
significantly better fit to the data. We find that the $P(N|M)$
parameter distributions are nearly  the same for the two models,
differing by, at most, $\sim 3\sigma$. Therefore, the improved fits for
the PNMCG model principally arise from the freedom to vary the density
profile of satellite galaxies. Varying the  density profile is thus
necessary to find a better fit to the data as we go to higher
luminosities.

We note that the reduced $\chi^{2}$ values (listed in each panel) are in
many cases quite high, even in the  PNMCG case.  This could mean that
the PNMCG model contains incorrect assumptions or does not have enough
freedom.  On the other hand, it could mean that the J11 jackknife
errors are underestimated.  To check the impact of the error estimates
on our modeling, we re-estimated errors for the $M_{r} < -20$ sample using
mock galaxy catalogs from the LasDamas project (McBride et al. in
prep.).  We used 160 catalogs
\footnote[1]{North-only SDSS footprint from the LasDamas "gamma" data
release.} and measured the dispersion  of $\wprp$ between the
catalogs, using the same binning method as J11.  We
then applied the fractional error (with respect to the mean of all
mock measurements) to the data (non-mock) measurement to estimate the
absolute errors and full covariance.  Finally, using our new mock
based  error estimates, we re-ran the MCMC chains for the PNMCG model.
We then compared the best-fit parameter values for the two fits and
found that the parameters did not change significantly.  By this, we
mean the difference in $\chi^{2}$ between the two best-fit points were
within $1\sigma$  when evaluated with either of the likelihood
surfaces (from each of the two MCMC runs with different errors).  We
conclude two things from this  test: (1) our somewhat high $\chi^{2}$
values are not overly concerning, and are likely due to a slight
underestimate  of the errors from jackknife re-sampling on the data,
and (2) this issue does not seem to affect any of our conclusions.


We now investigate the luminosity dependence of the radial
distribution of satellite galaxies and the degree to which it differs
from an NFW  distribution.  As discussed in \S\ref{theory}, when
constructing the  real-space correlation function in the halo model,
the one-halo term  considers contributions from central-satellite and
satellite-satellite  pairs.  The central-satellite contribution, which
is essentially just  the density profile itself (see Eq.~4 of W10), is
steeper than the  satellite-satellite pair contribution and thus
dominates the correlation  function on the very small scales that we
are considering  \citep[e.g., Figure 4. of][]{zheng09}.  Therefore,
the luminosity  dependence of the \emph{slope of $\xir$ on small
scales} can give a direct indication of the luminosity dependence of
the \emph{radial profile  of satellite galaxies} (though the slope of
$\xir$ will be less steep than the slope of the radial density
profile due to the dampening effect of the satellite-satellite term).
Figure~\ref{fig:xi_ALL} shows the residuals from an $r^{-2}$ power law
from the PNMCG best-fit models before converting to the projected
correlation functions shown in Figure~\ref{fig:wpgg_ALL}.   The
amplitudes of the curves have been arbitrarily staggered simply to
make the plot more clear.  The dotted lines highlight the cases of
$r^{-1}$ and $r^{-2}$ power laws.  The slope of $\xir$ is clearly a
strong function of luminosity, being close to -1 for low luminosity
galaxies and going more and more towards -2 for the $M_{r} < -21$
sample and even  steeper for LRGs. The W10 result for the steepness of
the slope of $\xir$ for LRGs on small scales was also found by
\citet{almeida08}.  Using the \citet{bower06} semi-analytic model
applied to the Millenium simulation \citep{springel_etal05}, they
found that the LRG real-space correlation function follows an $\sim
r^{-2.07}$ power law shape down to the $\sim10\hkpc $ scale.

We next wish to directly investigate the radial profiles of satellite
galaxies that are required by the data and compare them to the NFW
profile.  For each luminosity sample, we choose a halo mass equal to
the mean value of $\Mone$ in the PNMCG MCMC chain for that sample.  We
choose $\Mone$  because it represents the typical size halo that
contributes central-satellite pairs to $\xir$ (smaller halos have no
satellites and larger halos are rare).  Specifying the halo mass sets
the amplitude, virial radius, and dark matter concentration in
Equation~\ref{eq:profile_gamma}.  The radial profile then only depends
on the parameters $\fgal$ and $\gamma$.  We then take the full MCMC
chain for the PNMCG model and sort it by $\chisqr$  from lowest to
highest.  We then randomly draw 50 links from the top 95\%  of the
chain.  Each of these links has distinct values of $\fgal$ and
$\gamma$, which we insert into Equation~\ref{eq:profile_gamma} in
order to construct individual radial profiles.  The light grey curves
in each panel of Figure~\ref{fig:rad_dist} show these 50 profiles
(multiplied by $4\pi r^2dr$ to convert them from density into mass
profiles).  These curves thus span the $95\%$ confidence region
allowed by the data.   The green curve in each panel shows the radial
profile corresponding to  the best-fit PNMCG model.  To compare with
the NFW profile for dark matter,  we also plot the case $\fgal=1$,
$\gamma=1$, shown by the yellow bands.  We assume that the dark matter
profile of halos has a $20\%$ uncertainty, which we represent as the
thickness of the bands (see \S\ref{discussion} for  discussion of the
uncertainties in the dark matter distribution in halos).   For
reference, the scale radius for any particular sample occurs where the
dark matter radial profile ``turns over''.  Figure~\ref{fig:rad_dist}
shows, once again, that galaxies of increasing  luminosity deviate
more from an NFW profile.  In fact, the figure seems to suggest that
there is a transition point, somewhere between an absolute $r$-band
magnitude of -19.5 and -20, where the radial profile of satellite
galaxies goes from being consistent with NFW to completely
inconsistent.  For the highest luminosity samples, the profiles
approach a power-law shape.


\begin{figure}
\begin{center}
\includegraphics[width=.5\textwidth,height=.5\textheight,angle=0]{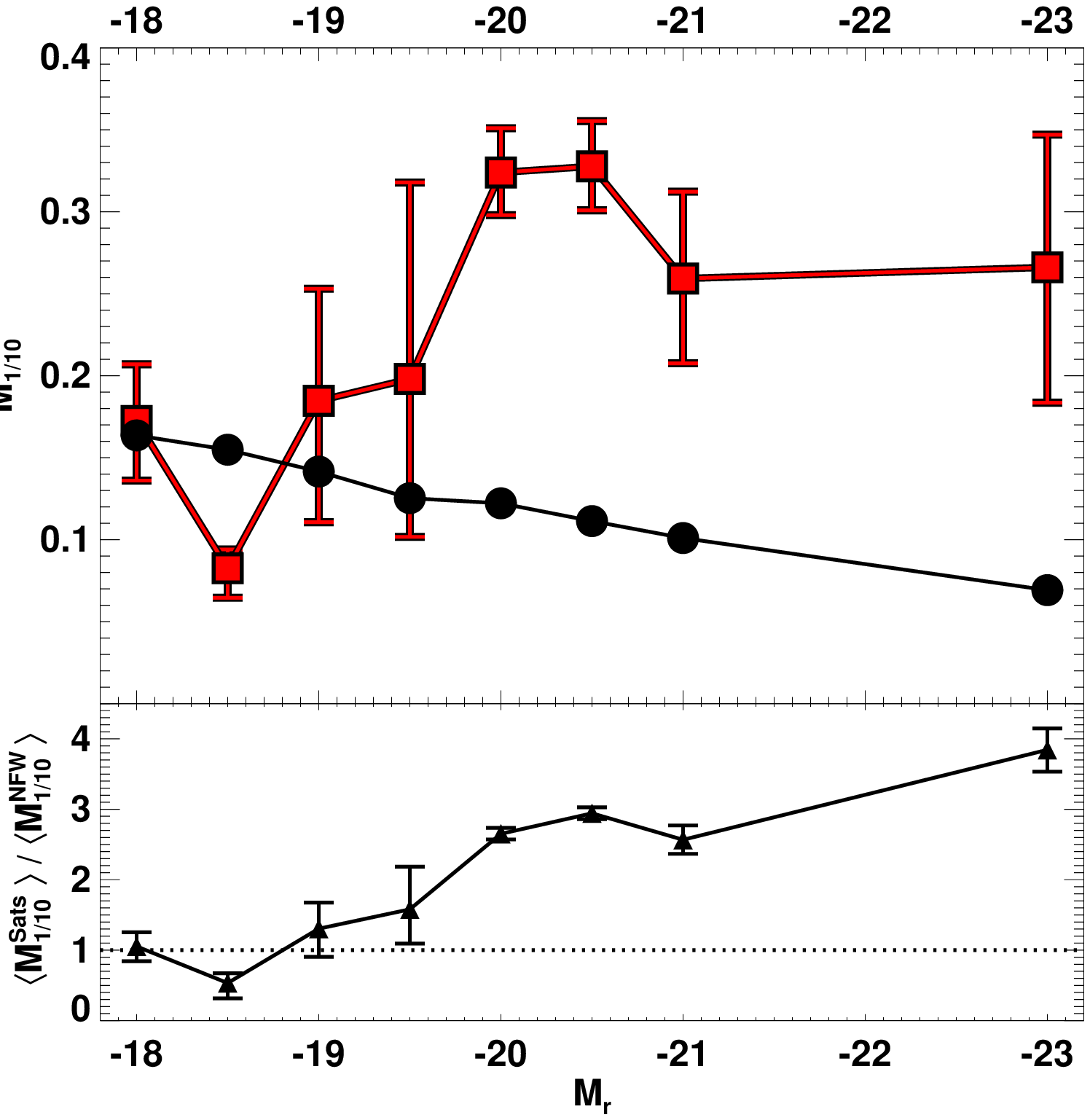}
\caption{A more physically useful definition of concentration: 
$M_{1/10}$, defined as the fraction of satellite galaxies (or mass) that are 
enclosed within one tenth of the virial radius of a halo (see Eq.~\ref{eq:M10}). 
\emph{Top Panel}: For each luminosity sample, we choose a mass equal to the 
mean value of $\Mone$ from the PNMCG MCMC chain, and we compute $M_{1/10}$
for each link in the chain.  Red squares (and connecting lines) show the mean 
of the $M_{1/10}$ distribution for a given luminosity, and errorbars for 
$M_{1/10}$ show the extrema of the middle $68.3\%$ of the distribution.  Black 
points show $M_{1/10}$ for an NFW distribution.
\emph{Bottom Panel}: Ratio of $M_{1/10}$ for satellites with respect to dark
matter.  The dotted line highlights a ratio of unity.  Satellite galaxies 
clearly become much more centrally concentrated than dark matter with 
increasing luminosity.}
\label{fig:Mr_vs_Mass}
\end{center}
\end{figure}


Figure \ref{fig:Mr_vs_Gamma} shows the luminosity dependence of the
inner  slope and concentration of satellite galaxies as found by the
PNMCG model.   The dark blue and light blue bands in each panel
represent the middle  $95\%$ and $68.3\%$ of the MCMC chains after
first sorting in $\gamma$  (top panel) and sorting in $\fgal$ (bottom
panel).  The filled red squares show the mean of the respective
parameters in the MCMC chain for each luminosity sample and filled
green squares show the best-fit values. The  solid black line
highlights the NFW dark matter inner slope of the density profile
which corresponds to $\gamma=\fgal=1$ and the dotted lines represent
the possible $20\%$ ambiguity of an assumed NFW inner slope and
concentration.  The $M_{r} < -18, -18.5, -19 \text{ and } -19.5$
samples have large spreads in $\gamma$ and $\fgal$ values, as there
are  many parameter combinations that can yield a similar
goodness-of-fit to  the data.   For $M_{r} < -20$ and brighter
galaxies, both $\gamma$ and  $\fgal$ have tighter distributions and
deviate from NFW.   Figure~\ref{fig:Mr_vs_Gamma} shows that there is a
strong luminosity  dependence of the satellite galaxy profile,
i.e. \emph{$\gamma$ is an  increasing function of luminosity, and
$\fgal$ is a decreasing one}.   However, notice the unique coupling of
$\fgal$ and $\gamma$ for the $-20.5$  case, where $\fgal$ seems to
deviate from the aforementioned trend.   $\gamma$ for $-20.5$
satellite galaxies also strays from the trend, but it  is still
significantly larger than NFW, and still lies within $1\sigma$ of the
$M_{r} < -20$ and $M_{r} < -21$ values ($\fgal$ is within $2\sigma$).
However, the reason for this outlier is unclear, though it may be due
to the fact that $\wprp$ has a strong feature at roughly the scale
radius of the typical host halo (see the $M_{r} < -20.5$ panel of
Fig. \ref{fig:wpgg_ALL}).  This may cause a unique interplay between
$\fgal$ and $\gamma$, resulting in a mild departure from the trends.
For $M_{r} < -20$ and brighter galaxies, the mean value of $\gamma$ ranges
between 1.6 - 2.1, significantly steeper than NFW.  This range of
$\gamma$ values persists all the way down to $M_{r} < -19$ when
considering the best-fit values.  As mentioned in \S\ref{data}, M06
found a systematic error in the manner that light in the overlapping
region of LRG pairs was being allocated in the SDSS pipeline.  This
caused a slight boost in the small-scale correlation function and was
corrected for by M06.  This effect is expected to rapidly diminish for
lower luminosity samples, because the intrinsic sizes of galaxies will
be smaller.  However, we apply the same LRG correction to the J11
$M_{r} < -21$ sample to model the maximum effect that this could have.
While we expect $\langle\gamma\rangle$ to decrease as a result of the
innermost data points being scaled to lower $\wprp$ values, we still
find that $\langle\gamma\rangle \sim 1.6$.  Thus, even after assuming
a \emph{drastic} over-correction, $\gamma$ is still strongly
discrepant from NFW.

Figure \ref{fig:Mr_vs_Gamma} showed the luminosity trends for $\gamma$
and $\fgal$ by sorting the MCMC chains separately for each parameter.
However, $\gamma$ and $\fgal$ are intrinsically coupled to one another
so it is important to investigate the joint $\gamma - \fgal$ parameter 
space.  Figure \ref{fig:gamma_fgal_space} Shows the $\gamma - \fgal$ 
parameter space at each luminosity, with the dark blue and light blue 
regions defined as the top $95\%$ and $68.3\%$ of the MCMC chain, after 
sorting in $\chisqr$.  There are two points of reference designated as 
filled circles: the $\gamma = 1, \fgal = 1$ combination for an 
NFW distribution and a $\gamma = 2, \fgal \rightarrow 0$ combination 
representing a singular isothermal sphere (SIS) distribution.  Red 
squares represent the mean values of $\gamma$ and $\fgal$ from our MCMC 
runs and green squares show the best-fit combination.  As expected, 
the $M_{r} < -18$ through $-19.5$ exhibit a broad range in 
$\gamma - \fgal$ combinations, with the NFW combination lying within 
the top 68.3\% region (with the exception of the -18.5 sample).  The 
$\gamma - \fgal$ regions clearly drift away from NFW for 
$M_{r} < -20$ and greater luminosities and are well described by an 
SIS distribution for both the $-21$ and LRG samples. 
Figure \ref{fig:gamma_fgal_space} shows that the inner slope and 
concentration parameters are strongly degenerate with each other,
particularly at low luminosities.  This can be caused by having
the scale radius (the radius within the halo where the slope
transitions from $-\gamma$ to -3) too close to the innermost data point.
When this happens, the data cannot accurately constrain the inner slope,
and only constrains $\gamma$ and $\fgal$ through their degenerate 
contribution to the amplitude of the density profile.  A simple 
calculation of the scale radius using the virial definition of a 
halo (see \S\ref{theory}) gives $r_{s} \sim 10 \hkpc$ for the 
$M_{r} < -18$ sample (adopting a halo mass equal to the mean value of 
$\Mone$), which is precisely at the scale of the innermost 
data point of $\wprp$. This leads to the large spread in $\gamma - \fgal$ 
combinations.  Of course, as we go to brighter samples, the scale radii
increase, and the parameters are better constrained.

For each luminosity, as the inner profile steepens, $\fgal$ decreases 
(with the slight deviation for the $M_{r} < -20.5$ case).  The decrease in 
concentration does not necessarily imply that the concentration of satellites 
is low in the traditional sense - i.e., that there are fewer galaxies at
small radii - but rather that the radial profiles simply evolve toward a more
power-law form.  For example, as $\fgal \rightarrow 0$ in the case of
LRGs, this does not mean that the LRG satellites are less concentrated
than the underlying dark matter.  As $\fgal \rightarrow 0$, the scale
radius is pushed far outside the virial radius, resulting in a pure 
power-law distribution of slope $-\gamma$.  Since the value of $\fgal$ 
can be misleading when thinking about the concentration of satellite 
galaxies, we consider an alternative definition of ``concentration'' that
is more physically useful: the fraction of satellite galaxies (or mass)
that are enclosed within a tenth of the virial radius $\Rvir$. We call
this quantity $M_{1/10}$ and compute it by integrating any radial profile out
to one-tenth the virial radius of the host halo and dividing by the total
mass out to $\Rvir$
\begin{equation}\label{eq:M10}
M_{1/10} = \frac{M(r < 0.1\Rvir)}{M(r < \Rvir)} .
\end{equation}

Figure \ref{fig:Mr_vs_Mass} (top panel) shows $M_{1/10}$ as a function
of luminosity.  For each luminosity sample, we assume a halo mass
equal to the mean value of $\Mone$ from the MCMC chain, and we
calculate $M_{1/10}$ for every link in the chain using each link's
values for $\gamma$ and $\fgal$.  We then find the mean of the
distribution of $M_{1/10}$ values (denoted by  the red squares and
connecting lines).  Errors for $M_{1/10}$ are given by  the extrema of
the middle $68.3\%$ of the distribution after sorting by $M_{1/10}$.
The black filled  circles show $M_{1/10}$ for a dark matter NFW
profile.  The dark matter concentration drops with luminosity because
the mean value of $\Mone$ increases with luminosity and concentration
is a decreasing function of halo mass \citep{bullock01}.  The bottom
panel of Figure~\ref{fig:Mr_vs_Mass}  shows the ratio of $M_{1/10}$
for satellites with respect to dark matter.  The figure clearly shows
that low luminosity satellite galaxies ($M_{r}>-19.5$) have spatial
concentrations similar to dark matter, whereas the profiles of
luminous satellite galaxies ($M_r<-20$ and brighter) are $\sim 2.5-4$
times more  concentrated than NFW.


\section{SUMMARY \& DISCUSSION}\label{discussion}


Do satellite galaxies trace the underlying dark matter distribution?
Our modeling of the small-scale $\wprp$ over an enormous range in
galaxy luminosity thresholds, from SDSS Main galaxies to LRGs ($M_{r} < -18$ 
through $ \lesssim -23$), has revealed a strong luminosity dependence of the 
radial distribution of satellite galaxies.  
We have found that for the low luminosity samples (lower than $M_{r} < -20$), 
there is a wide range of satellite galaxy concentrations (equal to $\fgal$ 
times the NFW dark matter concentration) and inner density profile slopes, 
$-\gamma$, that are consistent with the data.  This lack of constraining 
power is due to a strong degeneracy between $\fgal$ and $\gamma$, which, in
turn, is possibly due to the fact that the smallest scale data point in 
$\wprp$ is roughly around the scale radius for halos hosting galaxies of 
this size.  Nevertheless, low luminosity satellite galaxies have radial 
distributions that are generally consistent with the NFW dark matter distribution.
When we move to higher luminosity galaxies ($M_r<-20$ and brighter samples)
however, satellite profiles change dramatically, with the most striking
feature being that $\gamma$ jumps to much higher values.  Our results show 
that for $M_{r} < -20$ and brighter galaxies, $\gamma$ ranges from $\sim
1.6-2.1$, highly discrepant from NFW, even after assuming a $20\%$
inaccuracy in the dark matter profile.  Luminous satellite galaxies
are thus poor tracers of the underlying dark matter within halos on the small
scales that we probe. 
Since the effects of $\fgal$ and $\gamma$ on the radial profile are intertwined,
we have calculated a more physically useful quantity $M_{1/10}$, which 
gives the fractional amount of mass enclosed within one-tenth of the virial 
radius.  We find that $M_{1/10}$ is a strong function of luminosity, being
consistent with NFW for low luminosity galaxies and being $\sim 2.5-4$ times
more concentrated than NFW for galaxies brighter than $M_r<-20$.

Several other studies have also investigated the radial distribution
of galaxies within halos.  When considering faint satellites around
$\sim \Lstar$ galaxies, no consensus emerges.  \citet{chen09} found
the satellite distribution to be consistent with NFW.  \citet{more09b}
found a cored satellite profile best descibed by a $\gamma = 0, \fgal
=0.5$ combination, inconsistent with NFW. \citet{carlberg09} found
that the radial profiles of dwarf satellites around nearby galaxies
are much \emph{more} concentrated than NFW, rendering all of these
studies in disaccord. Furthermore, the discrepancies persist when
considering a broader range of galaxy systems. \citet{nierenberg11}
recently studied the radial profile of satellites around massive,
early-type galaxies at intermediate redshifts.  Assuming a power-law
profile model, they find that the satellites have an isothermal
distribution of slope -1.  \citet{guo_etal12} considered SDSS
satellite galaxies around host central galaxies of a wide luminosity
range and, in general, satellite number density profiles were shown to
be consistent with NFW.  The NFW dark matter concentration was found
to decrease with increasing satellite galaxy luminosity, independent
of the host galaxy luminosity (except for the brightest centrals).
They also detected a slightly \emph{steeper} profile for
\emph{fainter} satellites.  These varying results imply that
uncovering the spatial distribution of satellite galaxies is a
difficult problem.  In the end, we have found that there is a strong
luminosity dependence of the radial distribution of satellite
galaxies, wherein $\sim \Lstar$ and brighter satellites within group
and cluster sized halos have a \emph{substantially steeper} radial
profile than dark matter on scales smaller than $\sim 100\hkpc$.  We
emphasize that while the scales involved in all of these studies are
similar  to ours, galaxy sample selection is different in each case,
making it difficult to  directly compare their results with ours.


Our results have shown that luminous galaxies are poor tracers of
NFW.  We now address the possibility that NFW is a
poor model for the underlying dark matter distribution.  In the
established $\Lambda$CDM concordance cosmology, recent studies
invoking high resolution N-body simulations have shown that the mass
profiles of $\Lambda$CDM halos slightly, but systematically, deviate
from NFW, becoming shallower at smaller radii
\citep{stadel_etal09,delpopolo10}.  In fact, they may be better
described by Einasto profiles, which include a parameter $\alpha$ that
controls how the logarithmic slope will vary with radius to accurately
account for the fact that halo profiles seem to not be self-similar
\citep{gao08,navarro10,ludlow_etal11}.  These studies find, on
average, $\gamma < 1$ \citep{graham06,navarro10}.  However, on
average, the simpler,  two-parameter NFW model, which has a
characteristic $r^{-1}$ inner slope,  is accurate to within $10-20\%$
\citep{bensonReview10}.  Moreover,  \citet{mandelbaum08a} fit galaxy
cluster weak lensing profiles on small  scales and found that NFW and
Einasto profiles gave the same result to  within several percent.
Therefore, for simplicity, and to be consistent  with previous
modeling of LRGs by \citet{watson_etal10}, we assumed an  NFW
distribution throughout our modeling analysis.  We conclude from the
aforementioned studies that NFW is not a poor model for the dark
matter distribution in collisionless simulations and, to the extent
that it is, the true profile is even more discrepant from our results
for luminous satellite galaxies.

Of course, by assuming a pure dark matter profile established from
high resolution N-body simulations, the effects that baryons can have
on dark matter are not considered.  \emph{Adiabatic Contraction} (AC)
\citep{blumenthal86,ryden_gunn87,gnedin04} may cause a steepening of
the inner slope of the dark matter density profile as the gas
condenses and sinks to to the center of the dark matter potential well
\citep{diemand04,fukushige04,reed05,delpopolo09}.  However, the 
majority of results indicate that the steepening is insubstantial or
can vary widely from halo to halo \citep[e.g.,][]{tissera10}.  In fact,
while AC may cause a steepening of the inner profile, over time, major
and minor mergers can cause the dark matter profile to become shallower.
\citep{elzant01,romanodiaz08,romanodiaz09,johansson09}.  Observational
confirmations of AC are notoriously difficult.  In the case of galaxy
clusters, X-ray analysis by \citet{zappacosta06} showed that processes
during halo formation (e.g. gravitational heating from merger events)
counteract AC and the mass profiles were still well described by NFW.
\citet{mandelbaum06a} used galaxy-galaxy lensing to show that the mass
density profile of LRG clusters is consistent with NFW.  
Recently, \citet{shulz_etal10} studied the
profiles of a large sample of SDSS ellipticals and found that their
dynamical mass measurements suggest that the measured excess mass on
small scales \emph{may} be consistent with the AC hypothesis.
However, for bright galaxies, our results are still not reconcilable
with current AC models.We thus conclude that luminous satellite
galaxies are indeed poor tracers of the underlying dark matter
distribution, even accounting for the effects of baryons on the dark 
matter.

It is perhaps not surprising that galaxies do not behave like test
particles as they orbit within a dark matter potential well.  Being
massive and extended, they are subject to dynamical mechanisms that
would not affect test particles, namely dynamical friction and mass loss
due to the tidal field of their host halo.  In fact, these mechanisms
should affect large and massive objects more than small ones, which could
explain the luminosity trend that we observe.  Satellite galaxies are 
thought to reside within subhalos, so it makes more sense to compare
the radial distribution of galaxies to that of subhalos, rather than just 
dark matter.  While subhalos tend to have a less concentrated radial 
distribution than dark matter \citep{ghigna98,ghigna00,gao04a,delucia04,nagai05}, 
the subhalo distribution has yet to be extended down to the extreme 
small scales that we have probed.  High resolution N-body simulations 
are just now becoming available that allow for subhalos to be distinguished 
at the $10 \hkpc$ separation level \citep[e.g.,][]{klypin10a} for all of 
the luminosity samples that we have studied.  Of course, subhalos may not 
be perfect tracers of galaxies, but it will be interesting to see whether
the distribution of subhalos as a function of mass follows the same trend
that we have uncovered for satellite galaxies as a function of luminosity.
If the primary cause of this trend is dynamical in nature (e.g., dynamical
friction), it should also show up with subhalos.  Either way, such a
comparison will yield insight into the relation between subhalos and 
galaxies on very small scales, as well as the complicated non-linear 
processes occuring towards the centers of host halos, crucial for galaxy 
formation theory.

\section{ACKNOWLEDGEMENTS}\label{acknowledgements}

AAB is supported by Vanderbilt University and the Alfred P. Sloan
Foundation.  We thank Zheng Zheng for providing us with his best-fit
2-halo term for all the galaxy samples.


\bibliography{/data2/dwatson/lib/bibliography/citations.bib}

\begin{thebibliography}{69}
\expandafter\ifx\csname natexlab\endcsname\relax\def\natexlab#1{#1}\fi

\bibitem[{{Abazajian} {et~al.}(2009){Abazajian}, {Adelman-McCarthy},
  {Ag{\"u}eros}, {Allam}, {Allende Prieto}, {An}, {Anderson}, {Anderson},
  {Annis}, {Bahcall}, \& et~al.}]{DR7_09}
{Abazajian}, K.~N., {et~al.} 2009, \apjs, 182, 543

\bibitem[{{Almeida} {et~al.}(2008){Almeida}, {Baugh}, {Wake}, {Lacey},
  {Benson}, {Bower}, \& {Pimbblet}}]{almeida08}
{Almeida}, C., {Baugh}, C.~M., {Wake}, D.~A., {Lacey}, C.~G., {Benson}, A.~J.,
  {Bower}, R.~G., \& {Pimbblet}, K. 2008, \mnras, 386, 2145

\bibitem[{{Benson}(2010)}]{bensonReview10}
{Benson}, A.~J. 2010, \physrep, 495, 33

\bibitem[{{Berlind} \& {Weinberg}(2002)}]{berlind02}
{Berlind}, A.~A., \& {Weinberg}, D.~H. 2002, \apj, 575, 587

\bibitem[{{Berlind} {et~al.}(2003){Berlind}, {Weinberg}, {Benson}, {Baugh},
  {Cole}, {Dav{\'e}}, {Frenk}, {Jenkins}, {Katz}, \& {Lacey}}]{berlind03}
{Berlind}, A.~A., {et~al.} 2003, \apj, 593, 1

\bibitem[{{Blanton} {et~al.}(2003){Blanton}, {Hogg}, {Bahcall}, {Baldry},
  {Brinkmann}, {Csabai}, {Eisenstein}, {Fukugita}, {Gunn}, {Ivezi{\'c}},
  {Lamb}, {Lupton}, {Loveday}, {Munn}, {Nichol}, {Okamura}, {Schlegel},
  {Shimasaku}, {Strauss}, {Vogeley}, \& {Weinberg}}]{blanton03}
{Blanton}, M.~R., {et~al.} 2003, \apj, 594, 186

\bibitem[{{Blanton} {et~al.}(2005){Blanton}, {Schlegel}, {Strauss},
  {Brinkmann}, {Finkbeiner}, {Fukugita}, {Gunn}, {Hogg}, {Ivezi{\'c}}, {Knapp},
  {Lupton}, {Munn}, {Schneider}, {Tegmark}, \& {Zehavi}}]{VAGC_05}
------. 2005, \aj, 129, 2562

\bibitem[{{Blumenthal} {et~al.}(1986){Blumenthal}, {Faber}, {Flores}, \&
  {Primack}}]{blumenthal86}
{Blumenthal}, G.~R., {Faber}, S.~M., {Flores}, R., \& {Primack}, J.~R. 1986,
  \apj, 301, 27

\bibitem[{{Bower} {et~al.}(2006){Bower}, {Benson}, {Malbon}, {Helly}, {Frenk},
  {Baugh}, {Cole}, \& {Lacey}}]{bower06}
{Bower}, R.~G., {Benson}, A.~J., {Malbon}, R., {Helly}, J.~C., {Frenk}, C.~S.,
  {Baugh}, C.~M., {Cole}, S., \& {Lacey}, C.~G. 2006, \mnras, 370, 645

\bibitem[{{Bullock} {et~al.}(2001){Bullock}, {Kolatt}, {Sigad}, {Somerville},
  {Kravtsov}, {Klypin}, {Primack}, \& {Dekel}}]{bullock01}
{Bullock}, J.~S., {Kolatt}, T.~S., {Sigad}, Y., {Somerville}, R.~S.,
  {Kravtsov}, A.~V., {Klypin}, A.~A., {Primack}, J.~R., \& {Dekel}, A. 2001,
  \mnras, 321, 559

\bibitem[{{Carlberg} {et~al.}(2009){Carlberg}, {Sullivan}, \& {Le
  Borgne}}]{carlberg09}
{Carlberg}, R.~G., {Sullivan}, M., \& {Le Borgne}, D. 2009, \apj, 694, 1131

\bibitem[{{Chen}(2009)}]{chen09}
{Chen}, J. 2009, \aap, 494, 867

\bibitem[{{Conroy} {et~al.}(2006){Conroy}, {Wechsler}, \&
  {Kravtsov}}]{conroy06}
{Conroy}, C., {Wechsler}, R.~H., \& {Kravtsov}, A.~V. 2006, \apj, 647, 201

\bibitem[{{Cooray} \& {Sheth}(2002)}]{cooray02}
{Cooray}, A., \& {Sheth}, R. 2002, \physrep, 372, 1

\bibitem[{{De Lucia} {et~al.}(2004){De Lucia}, {Kauffmann}, {Springel},
  {White}, {Lanzoni}, {Stoehr}, {Tormen}, \& {Yoshida}}]{delucia04}
{De Lucia}, G., {Kauffmann}, G., {Springel}, V., {White}, S.~D.~M., {Lanzoni},
  B., {Stoehr}, F., {Tormen}, G., \& {Yoshida}, N. 2004, \mnras, 348, 333

\bibitem[{{Del Popolo}(2010)}]{delpopolo10}
{Del Popolo}, A. 2010, \mnras, 408, 1808

\bibitem[{{Del Popolo} \& {Kroupa}(2009)}]{delpopolo09}
{Del Popolo}, A., \& {Kroupa}, P. 2009, ArXiv e-prints

\bibitem[{{Diemand} {et~al.}(2004){Diemand}, {Moore}, \& {Stadel}}]{diemand04}
{Diemand}, J., {Moore}, B., \& {Stadel}, J. 2004, \mnras, 353, 624

\bibitem[{{Dunkley} {et~al.}(2005){Dunkley}, {Bucher}, {Ferreira}, {Moodley},
  \& {Skordis}}]{dunkley05}
{Dunkley}, J., {Bucher}, M., {Ferreira}, P.~G., {Moodley}, K., \& {Skordis}, C.
  2005, \mnras, 356, 925

\bibitem[{{El-Zant} {et~al.}(2001){El-Zant}, {Shlosman}, \&
  {Hoffman}}]{elzant01}
{El-Zant}, A., {Shlosman}, I., \& {Hoffman}, Y. 2001, \apj, 560, 636

\bibitem[{{Fukushige} {et~al.}(2004){Fukushige}, {Kawai}, \&
  {Makino}}]{fukushige04}
{Fukushige}, T., {Kawai}, A., \& {Makino}, J. 2004, \apj, 606, 625

\bibitem[{{Gao} {et~al.}(2004){Gao}, {De Lucia}, {White}, \&
  {Jenkins}}]{gao04a}
{Gao}, L., {De Lucia}, G., {White}, S.~D.~M., \& {Jenkins}, A. 2004, \mnras,
  352, L1

\bibitem[{{Gao} {et~al.}(2008){Gao}, {Navarro}, {Cole}, {Frenk}, {White},
  {Springel}, {Jenkins}, \& {Neto}}]{gao08}
{Gao}, L., {Navarro}, J.~F., {Cole}, S., {Frenk}, C.~S., {White}, S.~D.~M.,
  {Springel}, V., {Jenkins}, A., \& {Neto}, A.~F. 2008, \mnras, 387, 536

\bibitem[{{Ghigna} {et~al.}(1998){Ghigna}, {Moore}, {Governato}, {Lake},
  {Quinn}, \& {Stadel}}]{ghigna98}
{Ghigna}, S., {Moore}, B., {Governato}, F., {Lake}, G., {Quinn}, T., \&
  {Stadel}, J. 1998, \mnras, 300, 146

\bibitem[{{Ghigna} {et~al.}(2000){Ghigna}, {Moore}, {Governato}, {Lake},
  {Quinn}, \& {Stadel}}]{ghigna00}
------. 2000, \apj, 544, 616

\bibitem[{{Gnedin} {et~al.}(2004){Gnedin}, {Kravtsov}, {Klypin}, \&
  {Nagai}}]{gnedin04}
{Gnedin}, O.~Y., {Kravtsov}, A.~V., {Klypin}, A.~A., \& {Nagai}, D. 2004, \apj,
  616, 16

\bibitem[{{Graham} {et~al.}(2006){Graham}, {Merritt}, {Moore}, {Diemand}, \&
  {Terzi{\'c}}}]{graham06}
{Graham}, A.~W., {Merritt}, D., {Moore}, B., {Diemand}, J., \& {Terzi{\'c}}, B.
  2006, \aj, 132, 2701

\bibitem[{{Guo} {et~al.}(2012){Guo}, {Cole}, {Eke}, \& {Frenk}}]{guo_etal12}
{Guo}, Q., {Cole}, S., {Eke}, V., \& {Frenk}, C. 2012, ArXiv e-prints

\bibitem[{{Jiang} {et~al.}(2011){Jiang}, {Hogg}, \& {Blanton}}]{jiang_etal11a}
{Jiang}, T., {Hogg}, D.~W., \& {Blanton}, M.~R. 2011, ArXiv e-prints

\bibitem[{{Johansson} {et~al.}(2009){Johansson}, {Naab}, \&
  {Ostriker}}]{johansson09}
{Johansson}, P.~H., {Naab}, T., \& {Ostriker}, J.~P. 2009, \apjl, 697, L38

\bibitem[{{Klypin} {et~al.}(2010){Klypin}, {Trujillo-Gomez}, \&
  {Primack}}]{klypin10a}
{Klypin}, A., {Trujillo-Gomez}, S., \& {Primack}, J. 2010, ArXiv e-prints

\bibitem[{{Kravtsov} {et~al.}(2004){Kravtsov}, {Berlind}, {Wechsler}, {Klypin},
  {Gottl{\"o}ber}, {Allgood}, \& {Primack}}]{kravtsov04a}
{Kravtsov}, A.~V., {Berlind}, A.~A., {Wechsler}, R.~H., {Klypin}, A.~A.,
  {Gottl{\"o}ber}, S., {Allgood}, B., \& {Primack}, J.~R. 2004, \apj, 609, 35

\bibitem[{{Ludlow} {et~al.}(2010){Ludlow}, {Navarro}, {Springel},
  {Vogelsberger}, {Wang}, {White}, {Jenkins}, \& {Frenk}}]{ludlow_etal11}
{Ludlow}, A.~D., {Navarro}, J.~F., {Springel}, V., {Vogelsberger}, M., {Wang},
  J., {White}, S.~D.~M., {Jenkins}, A., \& {Frenk}, C.~S. 2010, \mnras, 406,
  137

\bibitem[{{Mandelbaum} {et~al.}(2006){Mandelbaum}, {Seljak}, {Cool}, {Blanton},
  {Hirata}, \& {Brinkmann}}]{mandelbaum06a}
{Mandelbaum}, R., {Seljak}, U., {Cool}, R.~J., {Blanton}, M., {Hirata}, C.~M.,
  \& {Brinkmann}, J. 2006, \mnras, 372, 758

\bibitem[{{Mandelbaum} {et~al.}(2008){Mandelbaum}, {Seljak}, \&
  {Hirata}}]{mandelbaum08a}
{Mandelbaum}, R., {Seljak}, U., \& {Hirata}, C.~M. 2008, J. Cosmology
  Astroparticle Phys., 8, 6

\bibitem[{{Masjedi} {et~al.}(2008){Masjedi}, {Hogg}, \& {Blanton}}]{masjedi08}
{Masjedi}, M., {Hogg}, D.~W., \& {Blanton}, M.~R. 2008, \apj, 679, 260

\bibitem[{{Masjedi} {et~al.}(2006){Masjedi}, {Hogg}, {Cool}, {Eisenstein},
  {Blanton}, {Zehavi}, {Berlind}, {Bell}, {Schneider}, {Warren}, \&
  {Brinkmann}}]{masjedi06a}
{Masjedi}, M., {et~al.} 2006, \apj, 644, 54

\bibitem[{{More} {et~al.}(2009){More}, {van den Bosch}, {Cacciato}, {Mo},
  {Yang}, \& {Li}}]{more09b}
{More}, S., {van den Bosch}, F.~C., {Cacciato}, M., {Mo}, H.~J., {Yang}, X., \&
  {Li}, R. 2009, \mnras, 392, 801

\bibitem[{{Nagai} \& {Kravtsov}(2005)}]{nagai05}
{Nagai}, D., \& {Kravtsov}, A.~V. 2005, \apj, 618, 557

\bibitem[{{Navarro} {et~al.}(1997){Navarro}, {Frenk}, \& {White}}]{nfw97}
{Navarro}, J.~F., {Frenk}, C.~S., \& {White}, S.~D.~M. 1997, \apj, 490, 493

\bibitem[{{Navarro} {et~al.}(2010){Navarro}, {Ludlow}, {Springel}, {Wang},
  {Vogelsberger}, {White}, {Jenkins}, {Frenk}, \& {Helmi}}]{navarro10}
{Navarro}, J.~F., {et~al.} 2010, \mnras, 402, 21

\bibitem[{{Nierenberg} {et~al.}(2011){Nierenberg}, {Auger}, {Treu}, {Marshall},
  \& {Fassnacht}}]{nierenberg11}
{Nierenberg}, A.~M., {Auger}, M.~W., {Treu}, T., {Marshall}, P.~J., \&
  {Fassnacht}, C.~D. 2011, \apj, 731, 44

\bibitem[{{Peacock} \& {Smith}(2000)}]{peacock00a}
{Peacock}, J.~A., \& {Smith}, R.~E. 2000, \mnras, 318, 1144

\bibitem[{{Reed} {et~al.}(2005){Reed}, {Governato}, {Verde}, {Gardner},
  {Quinn}, {Stadel}, {Merritt}, \& {Lake}}]{reed05}
{Reed}, D., {Governato}, F., {Verde}, L., {Gardner}, J., {Quinn}, T., {Stadel},
  J., {Merritt}, D., \& {Lake}, G. 2005, \mnras, 357, 82

\bibitem[{{Romano-D{\'{\i}}az} {et~al.}(2009){Romano-D{\'{\i}}az}, {Shlosman},
  {Heller}, \& {Hoffman}}]{romanodiaz09}
{Romano-D{\'{\i}}az}, E., {Shlosman}, I., {Heller}, C., \& {Hoffman}, Y. 2009,
  \apj, 702, 1250

\bibitem[{{Romano-D{\'{\i}}az} {et~al.}(2008){Romano-D{\'{\i}}az}, {Shlosman},
  {Hoffman}, \& {Heller}}]{romanodiaz08}
{Romano-D{\'{\i}}az}, E., {Shlosman}, I., {Hoffman}, Y., \& {Heller}, C. 2008,
  \apjl, 685, L105

\bibitem[{{Ryden} \& {Gunn}(1987)}]{ryden_gunn87}
{Ryden}, B.~S., \& {Gunn}, J.~E. 1987, \apj, 318, 15

\bibitem[{{Schulz} {et~al.}(2010){Schulz}, {Mandelbaum}, \&
  {Padmanabhan}}]{shulz_etal10}
{Schulz}, A.~E., {Mandelbaum}, R., \& {Padmanabhan}, N. 2010, \mnras, 408, 1463

\bibitem[{{Scoccimarro} {et~al.}(2001){Scoccimarro}, {Sheth}, {Hui}, \&
  {Jain}}]{scoccimarro01a}
{Scoccimarro}, R., {Sheth}, R.~K., {Hui}, L., \& {Jain}, B. 2001, \apj, 546, 20

\bibitem[{{Sheth} {et~al.}(2001){Sheth}, {Diaferio}, {Hui}, \&
  {Scoccimarro}}]{sheth01}
{Sheth}, R.~K., {Diaferio}, A., {Hui}, L., \& {Scoccimarro}, R. 2001, \mnras,
  326, 463

\bibitem[{{Springel} {et~al.}(2005){Springel}, {White}, {Jenkins}, {Frenk},
  {Yoshida}, {Gao}, {Navarro}, {Thacker}, {Croton}, {Helly}, {Peacock}, {Cole},
  {Thomas}, {Couchman}, {Evrard}, {Colberg}, \& {Pearce}}]{springel_etal05}
{Springel}, V., {et~al.} 2005, \nat, 435, 629

\bibitem[{{Stadel} {et~al.}(2009){Stadel}, {Potter}, {Moore}, {Diemand},
  {Madau}, {Zemp}, {Kuhlen}, \& {Quilis}}]{stadel_etal09}
{Stadel}, J., {Potter}, D., {Moore}, B., {Diemand}, J., {Madau}, P., {Zemp},
  M., {Kuhlen}, M., \& {Quilis}, V. 2009, \mnras, 398, L21

\bibitem[{{Strauss} {et~al.}(2002){Strauss}, {Weinberg}, {Lupton}, {Narayanan},
  {Annis}, {Bernardi}, {Blanton}, {Burles}, {Connolly}, {Dalcanton}, {Doi},
  {Eisenstein}, {Frieman}, {Fukugita}, {Gunn}, {Ivezi{\'c}}, {Kent}, {Kim},
  {Knapp}, {Kron}, {Munn}, {Newberg}, {Nichol}, {Okamura}, {Quinn}, {Richmond},
  {Schlegel}, {Shimasaku}, {SubbaRao}, {Szalay}, {Vanden Berk}, {Vogeley},
  {Yanny}, {Yasuda}, {York}, \& {Zehavi}}]{strauss02}
{Strauss}, M.~A., {et~al.} 2002, \aj, 124, 1810

\bibitem[{{Tinker} {et~al.}(2008){Tinker}, {Kravtsov}, {Klypin}, {Abazajian},
  {Warren}, {Yepes}, {Gottl{\"o}ber}, \& {Holz}}]{tinker08a}
{Tinker}, J., {Kravtsov}, A.~V., {Klypin}, A., {Abazajian}, K., {Warren}, M.,
  {Yepes}, G., {Gottl{\"o}ber}, S., \& {Holz}, D.~E. 2008, \apj, 688, 709

\bibitem[{{Tinker} {et~al.}(2005){Tinker}, {Weinberg}, {Zheng}, \&
  {Zehavi}}]{tinker05}
{Tinker}, J.~L., {Weinberg}, D.~H., {Zheng}, Z., \& {Zehavi}, I. 2005, \apj,
  631, 41

\bibitem[{{Tissera} {et~al.}(2010){Tissera}, {White}, {Pedrosa}, \&
  {Scannapieco}}]{tissera10}
{Tissera}, P.~B., {White}, S.~D.~M., {Pedrosa}, S., \& {Scannapieco}, C. 2010,
  \mnras, 406, 922

\bibitem[{{Warren} {et~al.}(2006){Warren}, {Abazajian}, {Holz}, \&
  {Teodoro}}]{warren06}
{Warren}, M.~S., {Abazajian}, K., {Holz}, D.~E., \& {Teodoro}, L. 2006, \apj,
  646, 881

\bibitem[{{Watson} {et~al.}(2010){Watson}, {Berlind}, {McBride}, \&
  {Masjedi}}]{watson_etal10}
{Watson}, D.~F., {Berlind}, A.~A., {McBride}, C.~K., \& {Masjedi}, M. 2010,
  \apj, 709, 115

\bibitem[{{Watson} {et~al.}(2011){Watson}, {Berlind}, \&
  {Zentner}}]{watson_powerlaw11}
{Watson}, D.~F., {Berlind}, A.~A., \& {Zentner}, A.~R. 2011, \apj, 738, 22

\bibitem[{{York} {et~al.}(2000)}]{york00a}
{York}, D.~G., {et~al.} 2000, \aj, 120, 1579

\bibitem[{{Zappacosta} {et~al.}(2006){Zappacosta}, {Buote}, {Gastaldello},
  {Humphrey}, {Bullock}, {Brighenti}, \& {Mathews}}]{zappacosta06}
{Zappacosta}, L., {Buote}, D.~A., {Gastaldello}, F., {Humphrey}, P.~J.,
  {Bullock}, J., {Brighenti}, F., \& {Mathews}, W. 2006, \apj, 650, 777

\bibitem[{{Zehavi} {et~al.}(2005{\natexlab{a}}){Zehavi}, {Eisenstein},
  {Nichol}, {Blanton}, {Hogg}, {Brinkmann}, {Loveday}, {Meiksin}, {Schneider},
  \& {Tegmark}}]{zehavi05b}
{Zehavi}, I., {et~al.} 2005{\natexlab{a}}, \apj, 621, 22

\bibitem[{{Zehavi} {et~al.}(2004){Zehavi}, {Weinberg}, {Zheng}, {Berlind},
  {Frieman}, {Scoccimarro}, {Sheth}, {Blanton}, {Tegmark}, {Mo}, {Bahcall},
  {Brinkmann}, {Burles}, {Csabai}, {Fukugita}, {Gunn}, {Lamb}, {Loveday},
  {Lupton}, {Meiksin}, {Munn}, {Nichol}, {Schlegel}, {Schneider}, {SubbaRao},
  {Szalay}, {Uomoto}, \& {York}}]{zehavi04a}
------. 2004, \apj, 608, 16

\bibitem[{{Zehavi} {et~al.}(2011){Zehavi}, {Zheng}, {Weinberg}, {Blanton},
  {Bahcall}, {Berlind}, {Brinkmann}, {Frieman}, {Gunn}, {Lupton}, {Nichol},
  {Percival}, {Schneider}, {Skibba}, {Strauss}, {Tegmark}, \&
  {York}}]{zehavi11}
------. 2011, \apj, 736, 59

\bibitem[{{Zehavi} {et~al.}(2005{\natexlab{b}}){Zehavi}, {Zheng}, {Weinberg},
  {Frieman}, {Berlind}, {Blanton}, {Scoccimarro}, {Sheth}, {Strauss}, {Kayo},
  {Suto}, {Fukugita}, {Nakamura}, {Bahcall}, {Brinkmann}, {Gunn}, {Hennessy},
  {Ivezi{\'c}}, {Knapp}, {Loveday}, {Meiksin}, {Schlegel}, {Schneider},
  {Szapudi}, {Tegmark}, {Vogeley}, \& {York}}]{zehavi05a}
------. 2005{\natexlab{b}}, \apj, 630, 1

\bibitem[{{Zheng}(2004)}]{zheng04a}
{Zheng}, Z. 2004, \apj, 610, 61

\bibitem[{{Zheng} {et~al.}(2005){Zheng}, {Berlind}, {Weinberg}, {Benson},
  {Baugh}, {Cole}, {Dav{\'e}}, {Frenk}, {Katz}, \& {Lacey}}]{zheng05}
{Zheng}, Z., {et~al.} 2005, \apj, 633, 791

\bibitem[{{Zheng} {et~al.}(2007){Zheng}, {Coil}, \& {Zehavi}}]{zheng07}
{Zheng}, Z., {Coil}, A.~L., \& {Zehavi}, I. 2007, \apj, 667, 760

\bibitem[{{Zheng} {et~al.}(2009){Zheng}, {Zehavi}, {Eisenstein}, {Weinberg}, \&
  {Jing}}]{zheng09}
{Zheng}, Z., {Zehavi}, I., {Eisenstein}, D.~J., {Weinberg}, D.~H., \& {Jing},
  Y.~P. 2009, \apj, 707, 554

\end{thebibliography}


\end{document}